\documentclass[preprint,prc,aps,nofootinbib,superscriptaddress,showpacs]{revtex4}
\usepackage{graphicx}
\usepackage{epsfig}
\usepackage{bm}
\usepackage{amssymb}

\newcommand{\lag}{\mathcal L}

\newcommand{\be}{\begin{eqnarray}}
\newcommand{\ee}{\end{eqnarray}}
\begin{document}

\title{Kaon Condensation in a Neutron Star under Strong Magnetic Fields by Using the Modified Quark-meson Coupling Model}

\author{C. Y. Ryu} \email{cyryu@skku.edu}
\affiliation{Department of Physics,
Soongsil University, Seoul 156-743, Korea}



\author{S. W. Hong} \email{swhong@skku.ac.kr}
\affiliation{Department of Physics and Department of Energy Science,
Sungkyunkwan University, Suwon 440-746, Korea}


\begin{abstract}
We have considered the antikaon condensation in a neutron star
in the presence of strong magnetic fields
by using the modified quark-meson coupling (MQMC) model.
The structure of the neutron star is investigated
with various magnetic fields and different kaon optical potentials,
and the effects of the magnetic fields for kaon condensation is discussed.
When employing strong magnetic fields inside a neutron star
with hyperons and kaon condensation,
the magnetic fields can cause the equation of state to be stiff;
thus, a large maximum mass of the neutron star can be obtained.



\end{abstract}

\maketitle


\section{Introduction}
Recently, a two-solar mass neutron star, $\sim 1.97M_\odot$,
was reported in Ref. \cite{demorest2010},
and some other observations \cite{Lattimer:2010uk}
suggest the possible existence of massive neutron stars.
On the other hand, exotic phases, such as the existence of hyperons,
meson condensation and quark matter,
can make the equation of state (EOS) of compact stars soft, giving rise to
lower masses of the star of about ($1.5 - 1.7$) $M_\odot$
although the details may depend on the models and parameters.
Thus, the recent observation in Ref. \cite{demorest2010} may rule out
the existence of exotic phases or
place a limit on the critical densities at which exotic phases appear.

However, previous theoretical studies
\cite{brod2000,cardall2001,Rabhi:2009ih,shen2009,Ryu:2010zzb}
suggest the possibility that massive neutron stars
with exotic phases may exist if we take into account strong magnetic fields.
For example, in Ref. \cite{Ryu:2010zzb},
the effect of strong magnetic fields on a neutron star was investigated
by using quantum hadrodynamics (QHD) with hyperons included.
As an extension of Ref. \cite{Ryu:2010zzb},
we further investigate the effect of strong magnetic fields
on neutron stars with hyperons and kaon condensation included
by using the modified quark-meson coupling (MQMC) model.

The quark-meson coupling (QMC) model was first proposed in Ref. \cite{guichon88},
and was later modified, for instance, in Ref. \cite{mqmc96} to recover
relativistic phenomena by introducing a density-dependent bag constant;
hence, the model is called the modified quark-meson coupling (MQMC) model.
Both the QMC and the MQMC models have been applied to various cases,
such as a nucleon \cite{Hong:2000tp}, finite nuclei \cite{Saito:1996sf},
nuclear matter \cite{mqmc96}, and neutron stars \cite{RHHK}.
(See Ref. \cite{Saito:2005rv} for a review.)
In the QMC and the MQMC models, a nucleon is treated as a MIT bag
in which quarks interact with each other
by exchanging $\sigma$, $\omega$ and $\rho$ mesons.
Thus, by using the MQMC model, neutron star matter can be described
in terms of the quark degrees of freedom.
In this work, we use the MQMC model and calculate
the structure of a neutron star by including hyperons and kaon condensation.
In particular, we investigate the possibility of exotic phases by
taking into account strong magnetic fields.

\section{Theory}

The Lagrangian density of the QMC and the MQMC models for dense matter
in the presence of strong magnetic fields,
which can be incorporated by introducing the vector potential $A^{\mu}$,
can be written in terms of octet baryons,
leptons, and five meson fields as follows:
\be \lag &=& \sum_b \bar \psi_b \Big [ i \gamma_\mu \partial^\mu
 - q_b \gamma_\mu A^\mu - M_b^*(\sigma, \sigma^*) - g_{\omega b} \gamma_\mu \omega^\mu
 - g_{\phi b} \gamma_\mu \phi^\mu  \nonumber \\
&& - g_{\rho b} \gamma_\mu \vec \tau \cdot \rho^\mu
 - \frac 12 \kappa_b \sigma_{\mu \nu} F^{\mu \nu}
\Big ] \psi_b  
 + \sum_l \bar \psi_l \left [  i \gamma_\mu \partial^\mu
 - q_l \gamma_\mu A^\mu - m_l \right ] \psi_l  \nonumber \\
&& + \frac 12 \partial_\mu \sigma \partial^\mu \sigma - \frac 12 m_\sigma^2 \sigma^2
+ \frac 12 \partial_\mu \sigma^* \partial^\mu \sigma^* - \frac 12 m_{\sigma^*}^2 {\sigma^*}^2
- \frac 14 W_{\mu \nu}W^{\mu \nu} \nonumber \\
&& + \frac 12 m_\omega^2 w_\mu w^\mu
- \frac 14 \Phi_{\mu \nu} \Phi^{\mu \nu} + \frac 12 m_\phi^2 \phi_\mu \phi^\mu
- \frac 14 R_{i \mu \nu} R_i^{\mu \nu} + \frac 12 m_\rho^2 \rho_\mu \rho^\mu
- \frac 14 F_{\mu \nu} F^{\mu \nu},
\ee
where $b$ and $l$ denote the octet baryons
($p, ~n, ~\Lambda, ~\Sigma^+, ~\Sigma^0, ~\Sigma^-, ~\Xi^0, ~\Xi^-$)
and the leptons ($e^-$ and $\mu^-$), respectively.
The $\sigma$, $\omega$ and $\rho$ meson fields describe
the interaction between the nucleons ($N-N$ interaction) and
that between the nucleons and the hyperons ($N-Y$ interaction).
The interaction between the hyperons ($Y-Y$) is assumed to be mediated
by $\sigma^*$ and $\phi$ meson fields.
$W_{\mu\nu}$, $R_{i \mu \nu}$, $\Phi_{\mu\nu}$, and
$F_{\mu\nu}$ represent the field tensors of $\omega$, $\rho$,
$\phi$ and photon fields, respectively.
In the presence of strong magnetic fields, the interaction of
anomalous magnetic moments (AMMs) of baryons with
the external magnetic field needs to be considered and can be
written in the form of $\kappa_b \sigma_{\mu\nu} F^{\mu\nu}$,
where $\sigma_{\mu \nu} = \frac i2 [\gamma_\mu, \gamma_\nu]$
and $\kappa_b$ is the strength of AMM of a baryon; i.e., $\kappa_p = 1.79 \mu_N$
for a proton and $\kappa_n = -1.91 \mu_N$ for a neutron,
$\mu_N$ being the nuclear magneton.
In the MQMC model, the effective mass of a baryon $b$ is obtained
by using the MIT bag model:
\be
M^*_b = \sqrt{E^2_b - \sum_q \left(\frac{x_q}{R_b} \right)^2} ~,~~~
E_b = \sum_q \frac{\Omega_q}{R_b} - \frac{Z_b}{R_b} + \frac{4\pi
R_b^3}{3} B_b(\sigma, \sigma^*) ~. \label{eq:bagery}
\ee
Here, the bag constant $B_b (\sigma, \sigma^*)$ has the following density dependence:
\be
B_b(\sigma, \sigma^*) = B_{b0} ~ \exp \left \{ -\frac{4}{M_b} \left [ g_\sigma^b
\sum_{q=u,d} n_q \sigma
 + g_{\sigma^*}^b \left ( 3 - \sum_{q=u,d} n_q \right ) \sigma^* \right ] \right \},
\label{eq:bagcon}
\ee
where $M_b$ is the mass of a baryon in vacuum and $n_q$ is the number of quarks.
The bag constant in vacuum $B_{b0}$ and a phenomenological constant $Z_b$ are
fitted to reproduce the mass of a free baryon at a bag radius, $R_b = 0.6$ fm,
and their values are taken from Ref. \cite{RHHK}.
The energy of a quark $\Omega_q$ is given in terms of
the momentum of a quark in a bag $x_q$, the bag radius $R_b$, and the
effective mass of a quark
$m_q^* = m_q - g_\sigma^q \sigma -g_{\sigma^*}^q \sigma^*$
as $\Omega_q = \sqrt{x_q^2 + (R_b m_q^*)^2}$.
The value of $x_q$ is determined by the
boundary condition on the bag surface, $ j_0(x_q) = \beta_q\, j_1(x_q)$.

The Lagrangian for kaons, which are treated as point particles, can be given by
\be
\lag &=& D^*_\mu \bar K D^\mu K - {m_K^*}^2 \bar K K,
\ee
where the covariant derivative $D_\mu$ is given by
\be
D_\mu = \partial_\mu + i q_K A_\mu + i g_{\omega K} \omega_\mu +
i g_{\phi K} \phi_\mu + i g_{\rho K}\tau_{iK}\rho^i_\mu
\ee
and the effective mass of a kaon can be given by
\be m_K^* = m_K - g_{\sigma K} \sigma - g_{\sigma^* K} \sigma^*. \ee
The equations for the meson fields are given by
\be
m_\sigma^2 \sigma &=& \sum_b g_{\sigma b} C_b (\sigma) \rho_s^b +
  g_{\sigma K} \frac{m_K^*}{\sqrt{{m_K^*}^2+|q_{K^-}|B}} \rho_{K^-} ,\nonumber \\
m_{\sigma^*}^2 \sigma^* &=&  \sum_b g_{\sigma^* b} C_b (\sigma^*) \rho_s^b +
  g_{\sigma^* K} \frac{m_K^*}{\sqrt{{m_K^*}^2+|q_{K^-}|B}} \rho_{K^-}, \nonumber \\
m_\omega^2 \omega_0 &=&  \sum_b g_{\omega b} \rho_v^b - g_{\omega K} \rho_{K^-}, \nonumber \\
m_\phi^2 \phi_0 &=&  \sum_b g_{\phi b} \rho_v^b + g_{\phi K} \rho_{K^-}, \nonumber \\
m_\rho^2 \rho_{30} &=&  \sum_b I_3^b g_{\rho b} \rho_v^b - \frac 12 g_{\rho K} \rho_{K^-},
\label{eq:mesons-kaon}
\ee
where $\rho_s$ and $\rho_v$ are, respectively, the scalar and the vector densities, and the details are given in Ref. \cite{shen2009}. The $C_b(\sigma)$ and the $C_b(\sigma^*)$
are determined from the relations $g_{\sigma b}C_b(\sigma) = - \partial M_b^* / \partial \sigma$ and $g_{\sigma^* b}C_b(\sigma^*) = - \partial M_b^* / \partial \sigma^*$.
The energy spectra of $K^-$ is given by
\be
\omega_{K^-} = \sqrt{{m_K^*}^2 + k_z^2 + (2n+1)|q_{K^-}|B}
  - g_{\omega K} \omega_0 + g_{\phi K} \phi - g_{\rho K} \rho_{03},
\ee
and the chemical potential for the s-wave kaon ($n = k_z = 0$) is
\be
\mu_K = \sqrt{{m_K^*}^2 + |q_{K^-}|B}
  - g_{\omega K} \omega_0 + g_{\phi K} \phi - g_{\rho K} \rho_{03}.
\label{eq:che-kaon}
\ee
The production of kaons takes place
when the chemical equilibrium condition $\mu_n - \mu_p = \mu_K$ is met.
The energy density of kaon condensation
is given by
\be \varepsilon_{K^-} = \rho_{K^-}~\sqrt{{m_K^*}^2 + |q_{K^-}|B},
\ee
while the condensed kaons do not contribute to the pressure because 
they are in s-wave states.

The Dirac equations for octet baryons and leptons in the mean field
approximation are given by \be \Big [ i \gamma_\mu
\partial^\mu &-& q_b \gamma_\mu A^\mu - M_b^*(\sigma, \sigma^*) -
g_{\omega b} \gamma^0 \omega_0
 - g_{\phi b} \gamma^0 \phi_0  \nonumber \\
&-& g_{\rho b} \gamma^0 \tau_3 \rho_{30}
 - \frac 12 \kappa_b \sigma_{\mu \nu} F^{\mu \nu} \Big ] \psi_b = 0, \ee
\be
(i \gamma_\mu \partial^\mu
 - q_l \gamma_\mu A^\mu - m_l ) \psi_l = 0,
\ee where $A_\mu = (0, ~0, ~xB, ~0)$ refers to the constant
magnetic field $B$ assumed to be along the $z$-axis. The
energy spectra of baryons and leptons are given by
\be E_b^C &=& \sqrt{ {k_{b, z}}^2 + \left ( \sqrt{{M_b^*}^2 + 2 \nu |q_b|
B} - s \kappa_b B \right )^2}
+ g_{\omega b}\omega_0 + g_{\phi b} \phi_0 + g_{\rho b} I_3^b \rho_{30}, \nonumber \\
E_b^N &=& \sqrt{ {k_{b, z}}^2 + \left ( \sqrt{{M_b^*}^2 + k_x^2 + k_y^2} - s \kappa_b B \right )^2}
+ g_{\omega b}\omega_0 + g_{\phi b} \phi_0 + g_{\rho b} I_3^b \rho_{30}, \nonumber \\
E_l &=& \sqrt{ {k_{l, z}}^2 + m_l^2 + 2 \nu |q_l| B},
\ee
where $E_b^C$ and $E_b^N$ represent energies of charged baryons
and neutral baryons, respectively. The Landau quantization of a
charged particle due to magnetic fields is denoted as $\nu = n
+ 1/2 - sgn(q) s/2 = 0, 1, 2, \cdots~$, where $q$ denotes the electric charge and
$s=1(-1)$ represents the spin up (down) state.
The chemical potentials of baryons and
leptons are, respectively, given by \be \mu_b &=& E_f^b +
g_{\omega b}\omega_0 + g_{\phi b} \phi_0 + g_{\rho b} I_z^b
\rho_{30}, \label{eq:che-b}
\\
\mu_l &=& \sqrt{k_f^2 + m_l^2 + 2 \nu |q_l| B}, \label{eq:che-l}
\ee
where $E_f^b$ is the Fermi energy of a baryon $b$ and $k_f$ is the
Fermi momentum of a lepton $l$. For charged particles, $E_f^b$ is written as
\be {E_f^b}^2 = {k_f^b}^2 + (\sqrt{{M_b^*}^2 + 2 \nu |q_b| B} - s
\kappa_b B)^2,
\ee
where $k_f^b$ is the Fermi momentum of the baryon $b$. Since the Landau
quantization does not appear for neutral baryons, the Fermi energy
is simply given by
\be {E_f^b}^2 = {k_f^b}^2 + ({M_b^*} - s \kappa_b B)^2.
\ee

We impose the usual three constraints for neutron star matter:
baryon number conservation, charge neutrality, and
chemical equilibrium. The equations for the meson fields in Eq.
(\ref{eq:mesons-kaon}) are solved together with the chemical potentials of
baryons and leptons under the above three constraints. The total
energy density is given by $\varepsilon_{tot} = \varepsilon_m +
\varepsilon_f$, where the energy density for matter fields is given by
\be \varepsilon_m &=& \sum_b \varepsilon_b + \sum_l \varepsilon_l
+ \frac 12 m_\sigma^2 \sigma^2 + \frac 12 m_{\sigma^*}^2
{\sigma^*}^2 + \frac 12 m_\omega^2 \omega^2 + \frac 12 m_\phi^2
\phi^2 + \frac 12 m_\rho^2 \rho^2,
\ee
and the energy density due to the magnetic field is given by
$\varepsilon_f = B^2 / 2$. The total pressure can also be written
as \be P_{tot} = P_m + \frac 12 B^2, \ee where the pressure due to
matter fields is obtained by using $P_m = \sum_i \mu_i \rho_v^i -
\varepsilon_m$. The relation between the mass and the radius for a static
and spherically-symmetric neutron star is generated by solving
the Tolman-Oppenheimer-Volkoff (TOV) equations with the equation
of state (EOS) given above.

\section{Results}
\begin{figure}
\centering
\includegraphics[width=7.5cm]{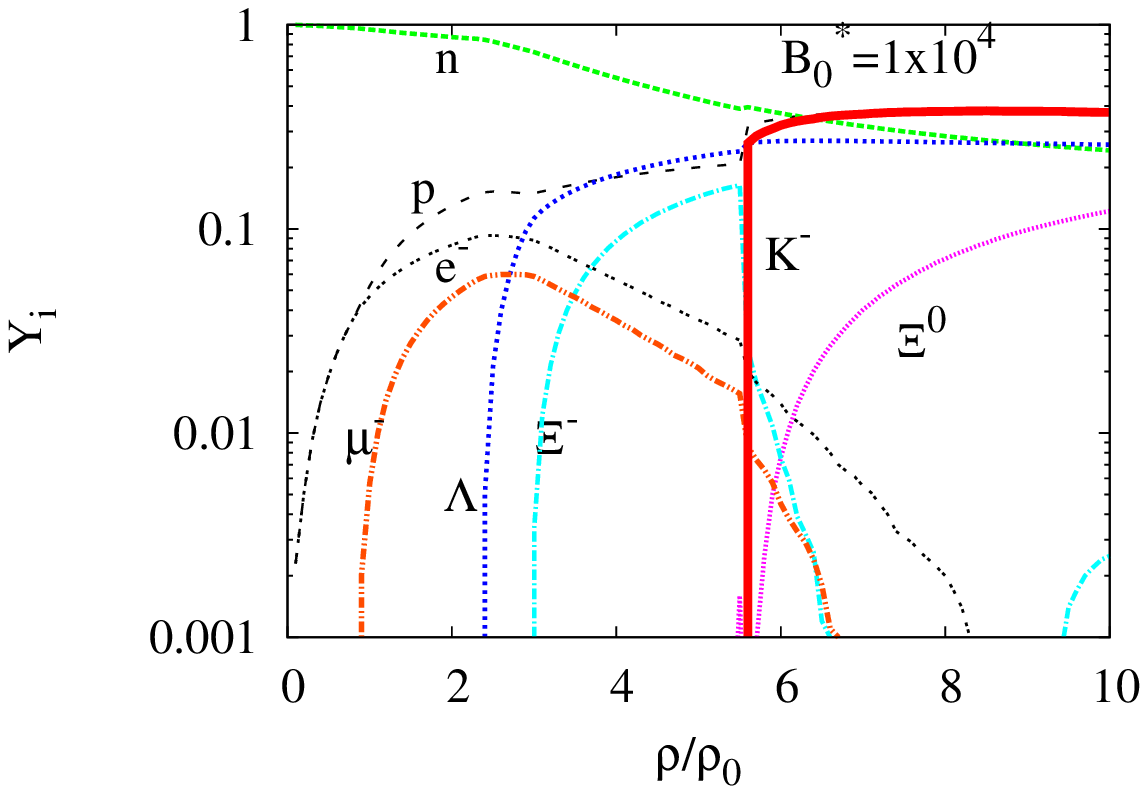}
\includegraphics[width=7.5cm]{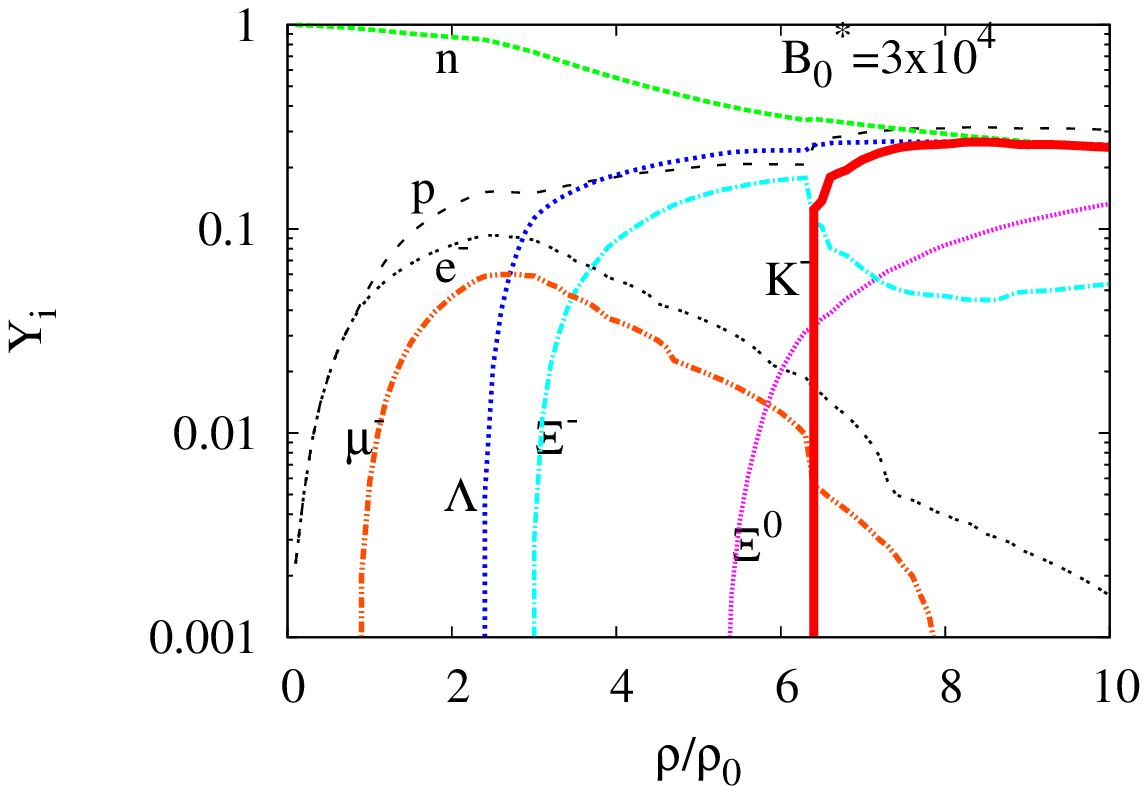} \\
\includegraphics[width=7.5cm]{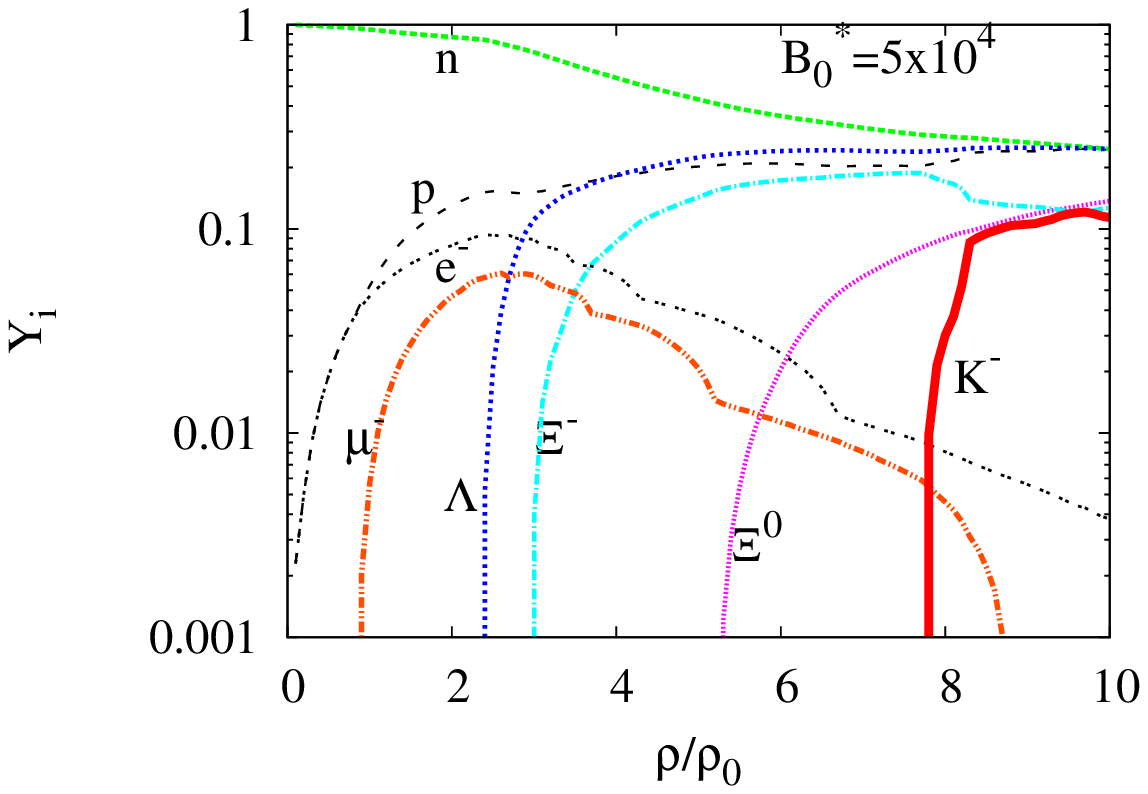}
\includegraphics[width=7.5cm]{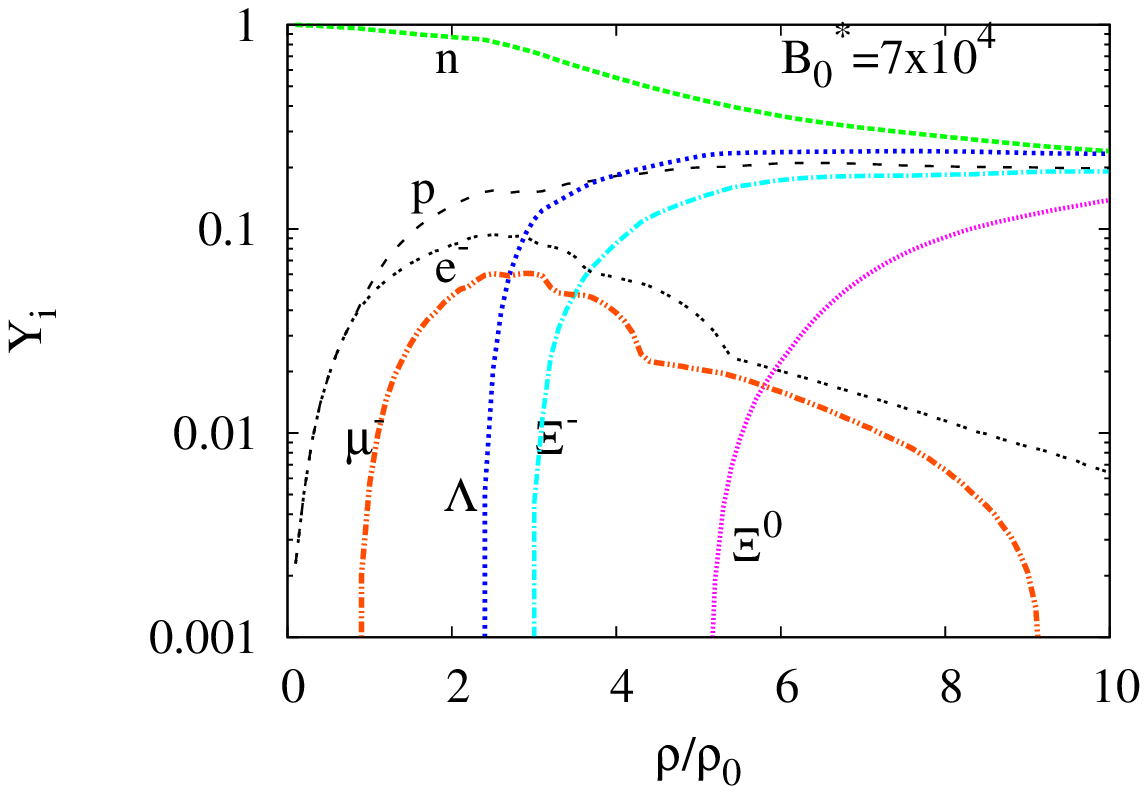}
\caption{Populations of particles, $Y_i$, in a neutron star for $U_K =  - 120$ MeV.}
\label{fig:popul120}
\end{figure}
\begin{figure}
\centering
\includegraphics[width=7.5cm]{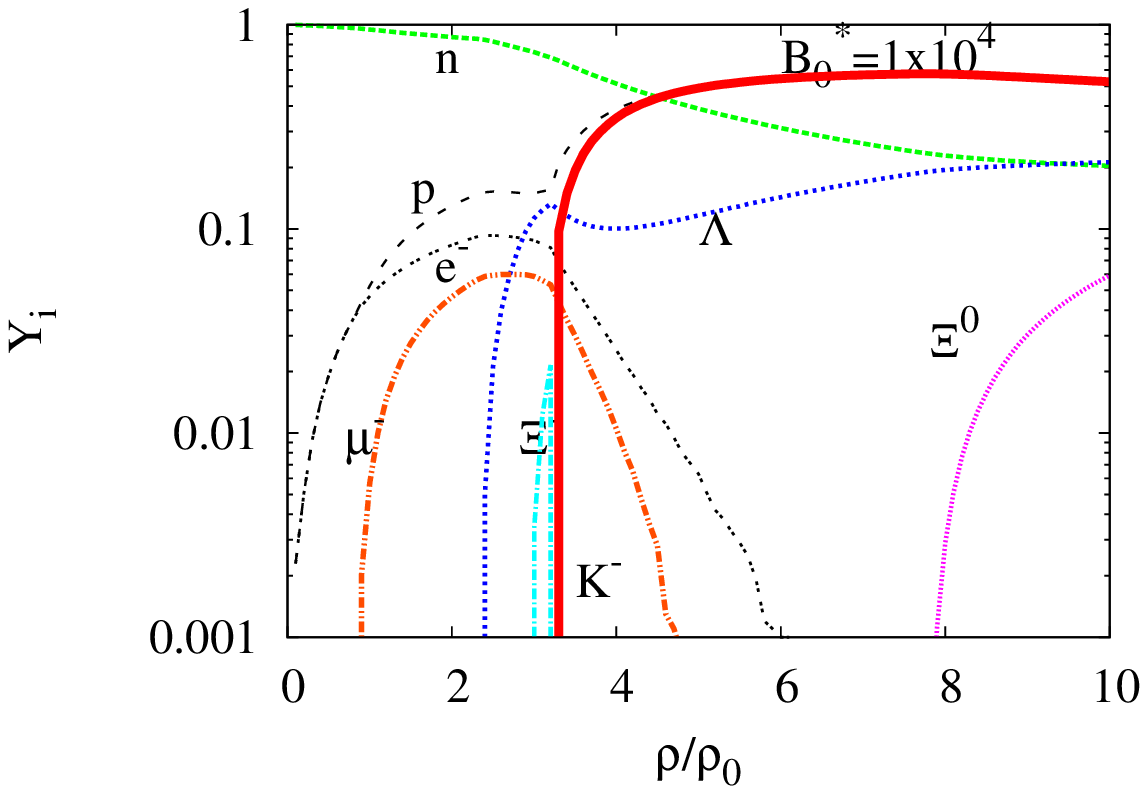}
\includegraphics[width=7.5cm]{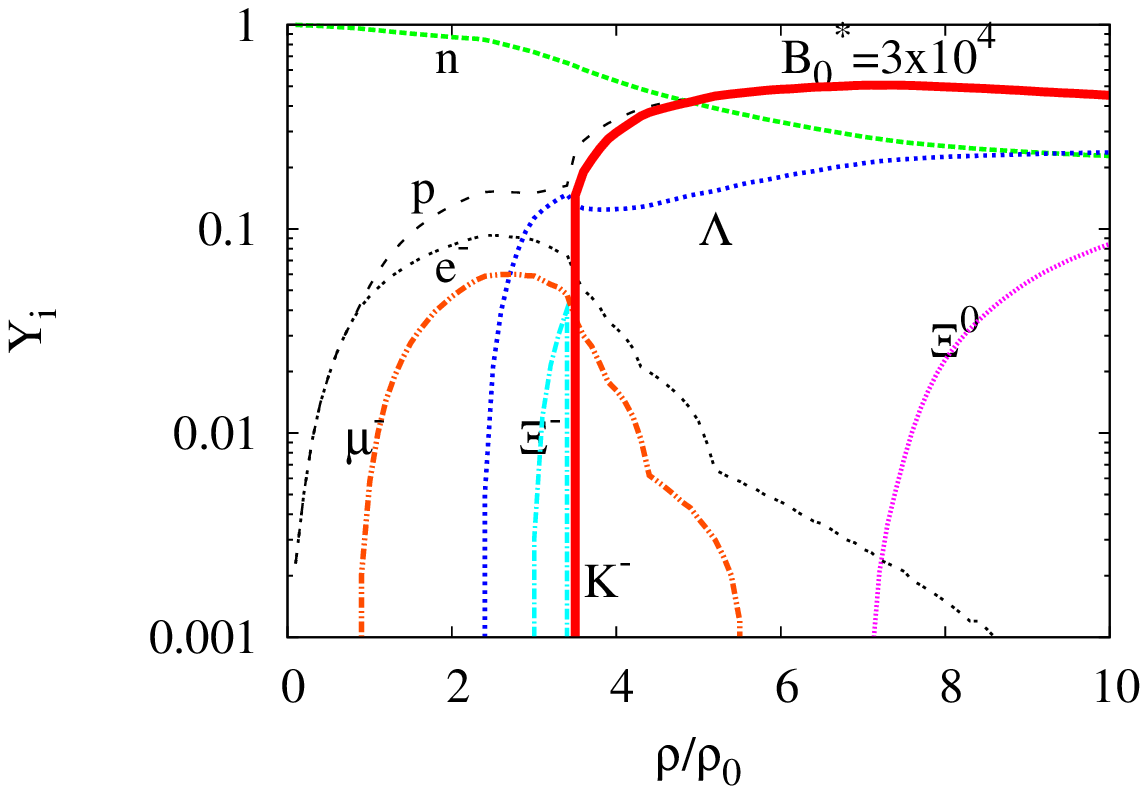} \\
\includegraphics[width=7.5cm]{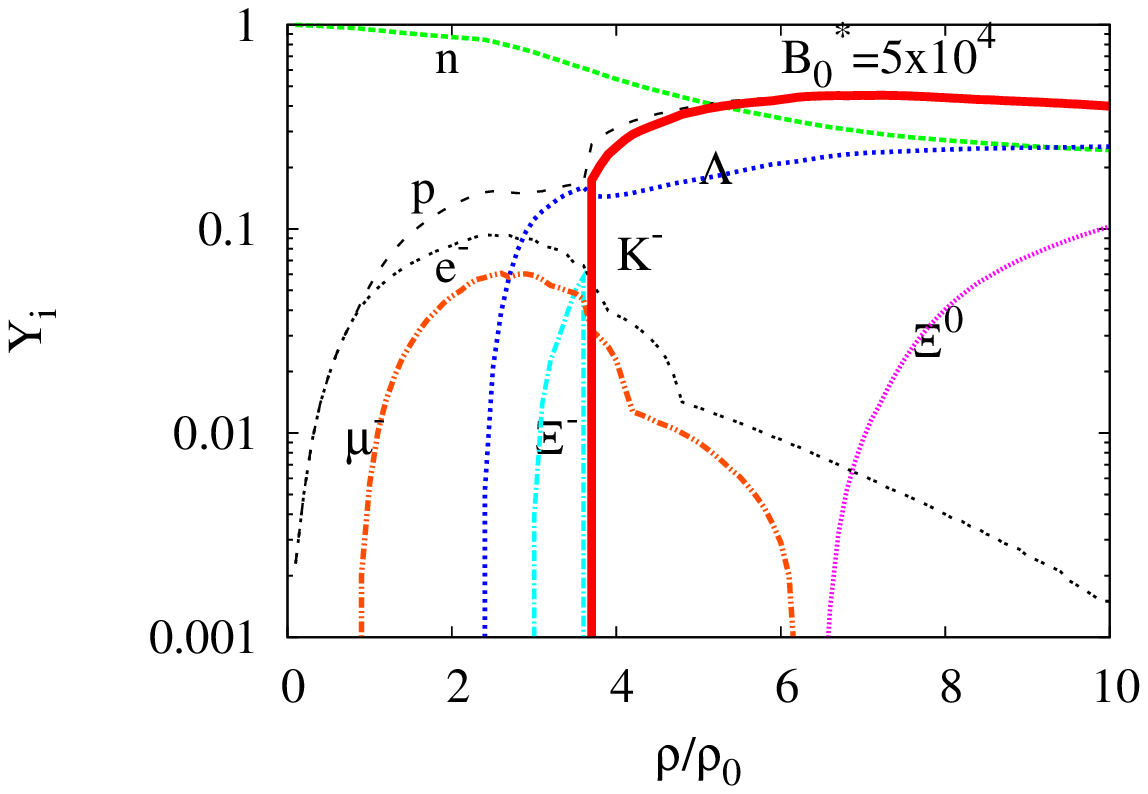}
\includegraphics[width=7.5cm]{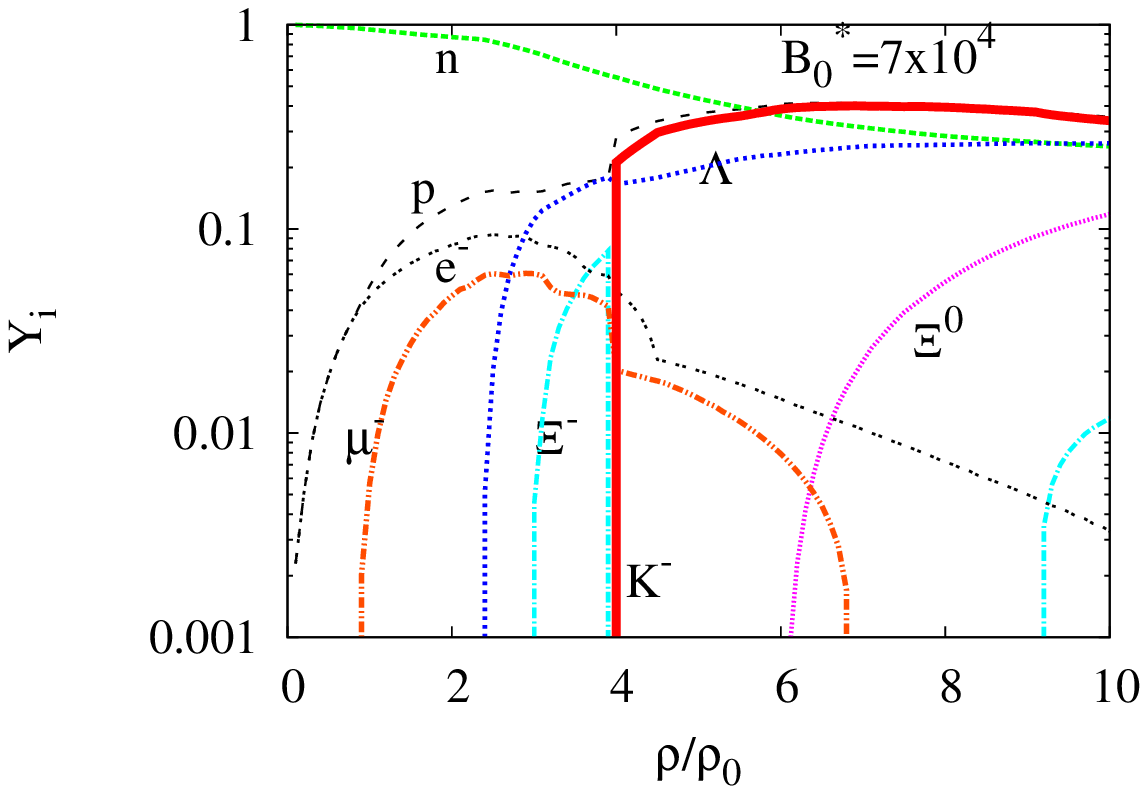}
\caption{Populations of particles, $Y_i$, in a neutron star for $U_K = - 140$ MeV.}
\label{fig:popul140}
\end{figure}
\begin{figure}
\centering
\includegraphics[width=7.5cm]{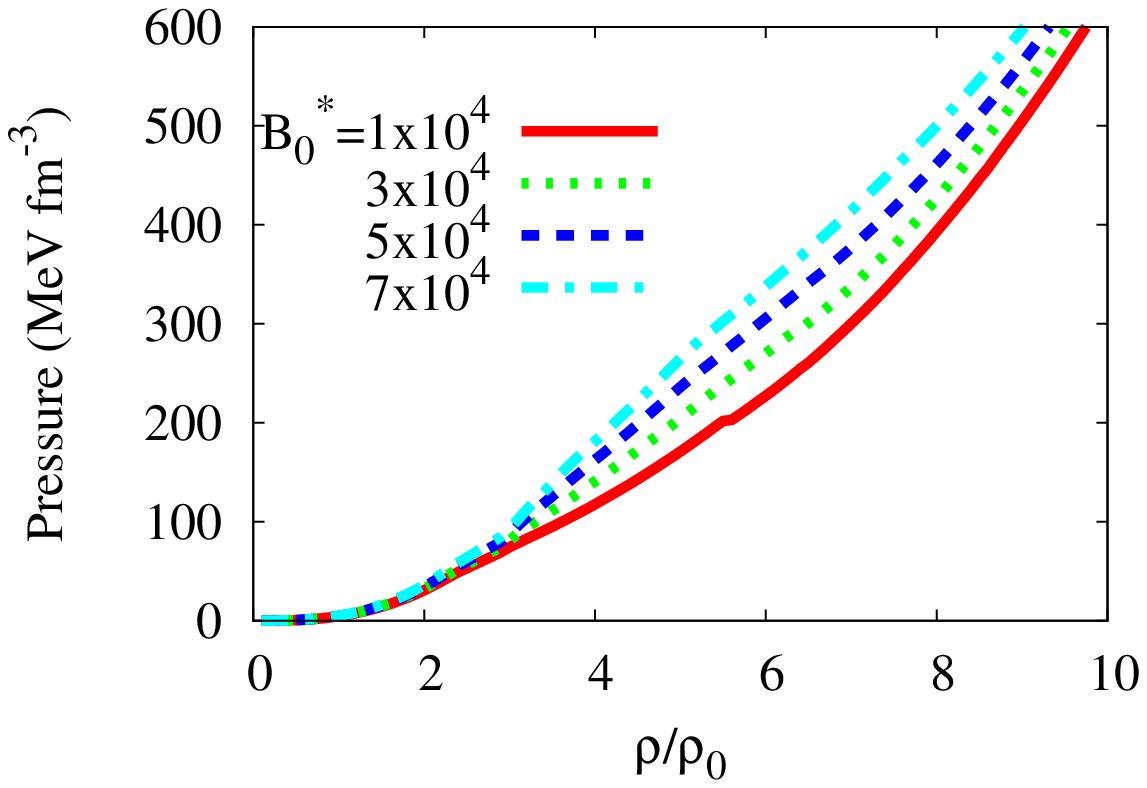}
\includegraphics[width=7.5cm]{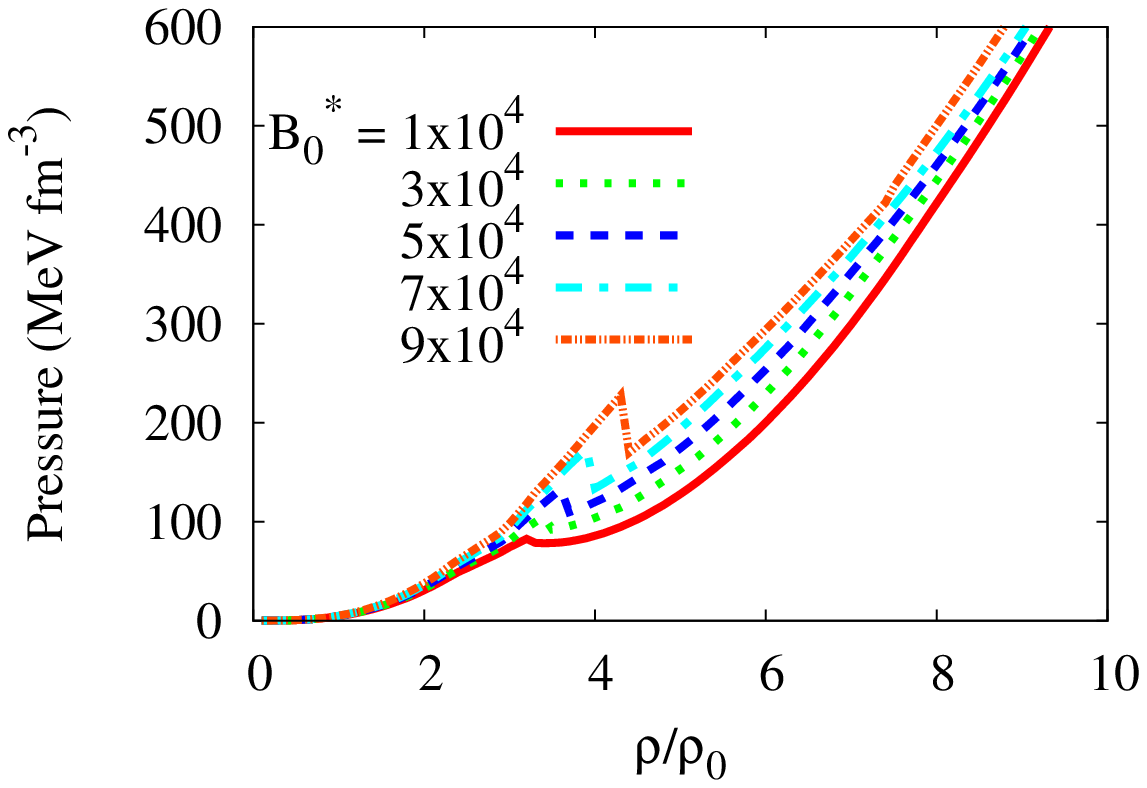}
\caption{Equation of state for $U_K =  -120$ (left panel) and
$- 140$ (right panel) MeV in various magnetic fields.}
\label{fig:eos}
\end{figure}
\begin{figure}
\centering
\includegraphics[width=7.5cm]{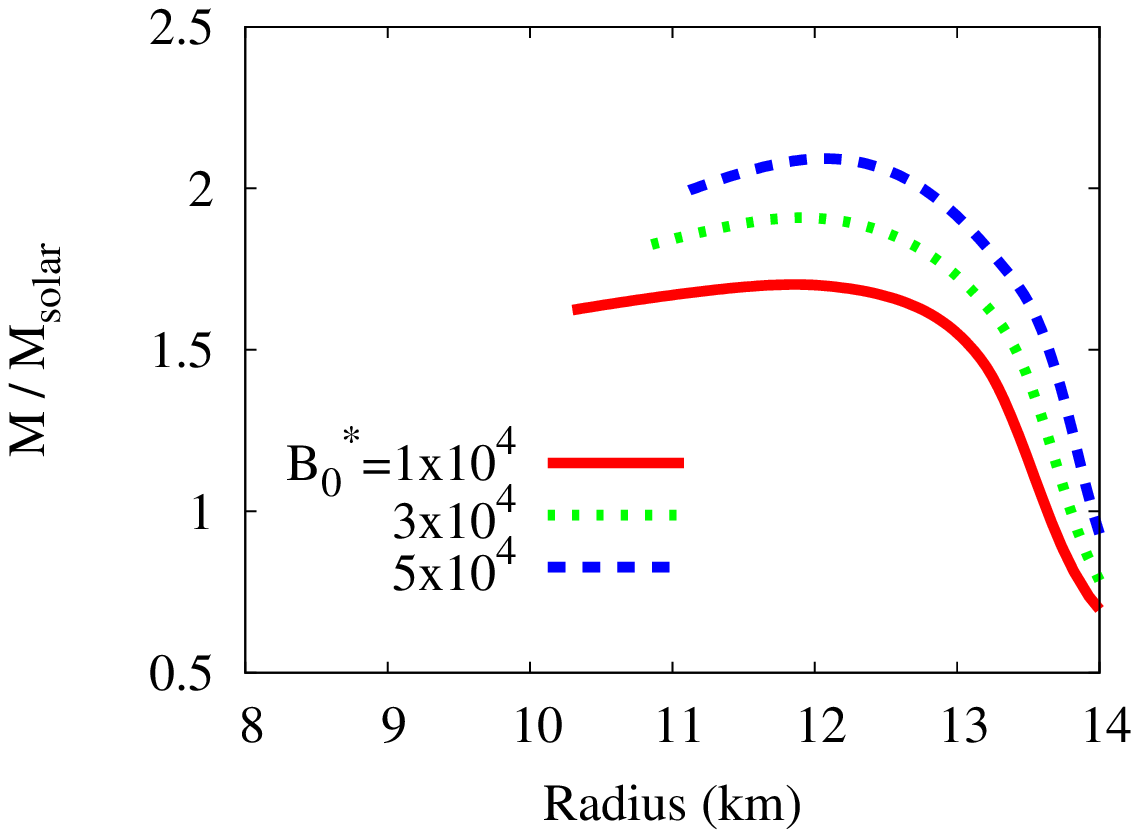}
\includegraphics[width=7.5cm]{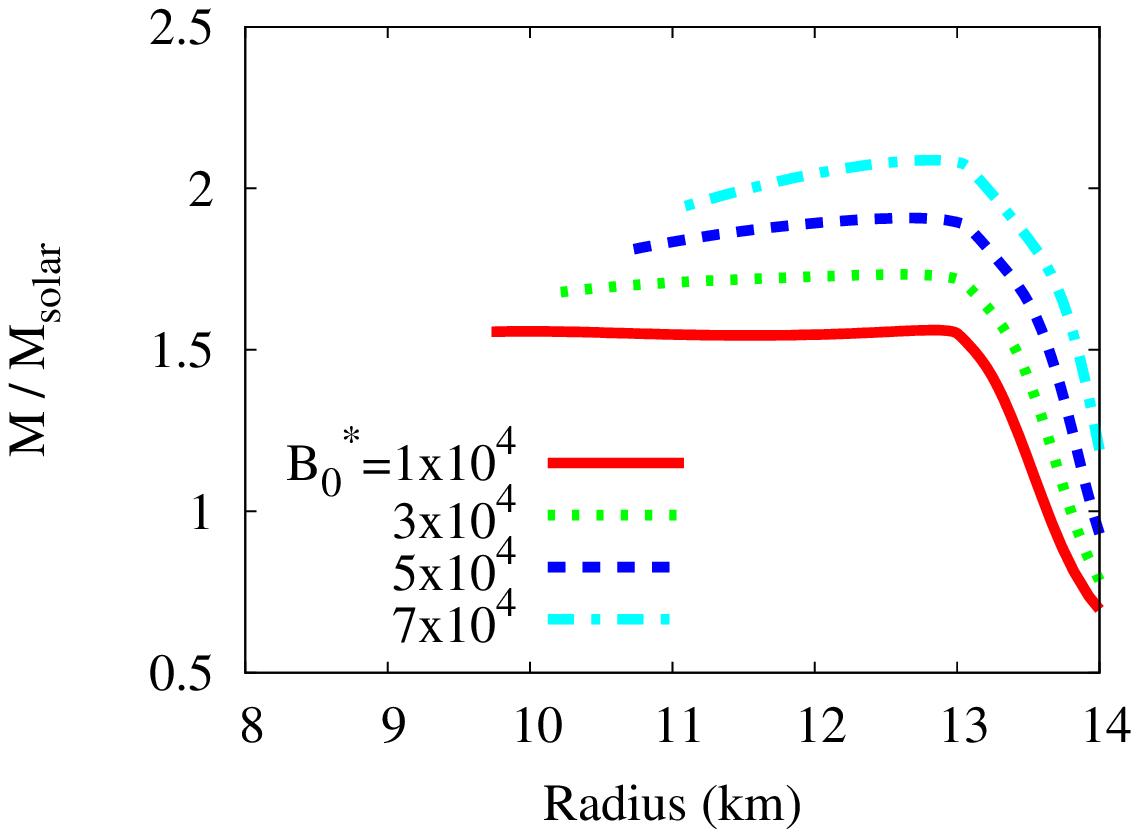}
\caption{Mass and radius relations
for $U_K =  -120$ (left panel) and $- 140$ (right panel) MeV
with various magnetic fields.}
\label{fig:mr}
\end{figure}

We use the parameter set in Ref. \cite{RHHK} for the coupling
constants between quarks and $\sigma$, $\omega$ and $\rho$ meson fields
and for the couplings between kaons and meson fields.
For the coupling constants of hyperons in
nuclear medium, $g_{\omega Y}$ and $g_{\phi Y}$ are determined by using the quark counting
rule and SU(6) symmetry \cite{RHHK}.
$g_\sigma^Y$ in Eq. (\ref{eq:bagcon}) is fitted to reproduce the potential of
each hyperon at saturation density, whose strengths are given by
$U_\Lambda = -30$ MeV, $U_\Sigma = 30$ MeV and $U_\Xi = -15$ MeV,
where the optical potential of a baryon is defined as $U_b = -(M_b - M_b^*) + g_{\omega b}\omega_0 $ at saturation density
($\rho_0 = 0.17$ fm$^{-3}$).
Since the magnetic fields may also depend on
density, we take the density-dependent magnetic fields used in
Refs. \cite{Rabhi:2009ih} and \cite{Ryu:2010zzb}:
\be B \left ( \rho / \rho_0 \right ) = B^{surf} + B_0 \left [ 1 -
\exp \{-\beta \left ( \rho / \rho_0 \right )^\gamma \} \right ],
\label{eq-B} \ee
where $B^{surf}$ is the magnetic field at the surface of a neutron
star, which is taken as $10^{15}$ G for this work, and $B_0$
represents the magnetic field saturated at high densities.

In the present work, we use the values $\beta = 0.02$ and $\gamma = 3$
in Eq. (\ref{eq-B}) \cite{Ryu:2010zzb}.
Since the magnetic field is usually
written in a unit of the critical field for the electron, $B_e^c =
4.414 \times 10^{13}$ G, the $B$ and the $B_0$ in Eq. (\ref{eq-B})
can be written as $B^* = B / B_e^c$ and $B_0^* = B_0 /B_e^c$.
Here, we treat the $B_0^*$ as a free parameter.

The populations of particles are shown with increasing strength of $B_0^*$ for
the kaon optical potentials $U_K = -120$ and $U_K = -140$ MeV
in Figs. \ref{fig:popul120} and \ref{fig:popul140}, respectively.
In Eq. (\ref{eq:che-kaon}), the chemical potential of a kaon, $\mu_K$,
is composed of $\sqrt{{m_K^*}^2 + |q_K|B}$ and the contribution from vector meson fields.
Thus, $\mu_K$ depends on the effective mass of the kaon,
the optical potential of kaons ($U_K = - g_{\sigma K}\sigma - g_{\omega K}\omega$)
and the strength of the magnetic field.
The condition for kaon condensation is $\mu_n - \mu_p = \mu_K$
so that electrons are replaced by kaons in beta equilibrium, which 
means that if $\mu_K$ is larger than $\mu_e$, kaon condensation does not take place.
Therefore, a lower $\mu_K$ makes the matter kaon dominant.
On the contrary, if $\mu_K$ has a higher value,
kaons are suppressed by negatively-charged particles, which are mainly electrons.
Since strong magnetic fields will increase
$\mu_K$, kaons are suppressed by strong magnetic fields.
Such phenomena are shown to appear in Figs. \ref{fig:popul120} and \ref{fig:popul140}.
As shown in Fig. \ref{fig:popul120}, for $U_K=-120$ MeV, kaons are totally suppressed
for $B_0^* = 7 \times 10^4$, and the neutron star matter becomes just hyperonic matter.
For $U_K=-140$ MeV, as shown in Fig. \ref{fig:popul140},
kaons still survive with $B_0^* = 7 \times 10^4$, which corresponds to
$B \sim 3 \times 10^{18}$ G.

In Fig. \ref{fig:eos}, the EOSs for various values of $B_0^*$
are shown for $U_K = -120$ and $-140$ MeV.
The EOS for $U_K = -120$ MeV shows a very smooth behavior
so that the phase transition seems to be second order.
In particular, the contribution from kaon condensation
is not significant with increasing magnetic fields;
thus, the EOS does not change dramatically.
However, the EOS for $U_K = -140$ MeV in the right panel
suggests a first-order phase transition,
having less suppression of kaons with increasing magnetic field.

The mass and the radius relations can be obtained
by solving the TOV equations. For $U_K = -140$ MeV,
we need to use the Maxwell condition.
The results for the mass and radius relations for various $B_0^*$ are shown
in Fig. \ref{fig:mr}.
In the case of $U_K = -120$ MeV,
the relations look similar to those of the hyperonic matter obtained
in Ref. \cite{Ryu:2010zzb}
because the contribution from kaon condensation is relatively small.
For $U_K = -140$ MeV, the maximum masses are less than those from $U_K = -120$ MeV
because of the effect of kaon condensation.
As a result, even with hyperons and kaon condensation, strong magnetic fields ($B \sim 2 \times 10^{18}$ G)
can make the mass as large as $\sim 2 M_\odot$.

\section{Summary}
We have investigated the behavior of kaon condensation
for various magnetic field strengths with large optical potentials of a kaon.
By considering strong magnetic fields inside a neutron star,
a large mass $ \sim 2 M_\odot$ can be obtained
even with hyperons and kaon condensation
although the results depend on the kaon optical potential.
In our model, if the kaon optical potential is smaller than $U_K = -120$ MeV,
the contribution from the kaon condensation is hardly seen.
Such a deep kaonic optical potential $U_K \sim -120$ MeV
at saturation density was predicted in Ref. \cite{Akaishi:2002bg}.
In the case of $U_K = -140$ MeV, the EOS shows a first-order phase transition.
We, thus, use the Maxwell condition to treat the first-order phase transition
and to calculate the mass and radius relations of the neutron star from the TOV equations.
We find that $B \sim 2 \times 10^{18}$ G is needed to explain $2 M_\odot$,
though we do not mean that the strong magnetic field we considered in this work
can explain the mass of PSR J1614-2230 of Ref.~[1].
The massive neutron star observed in Ref.~[1]
is in a neutron star-white dwarf binary system and is unlikely to have
a strong magnetic field.
The magnetic field on the surface of PSR J1614-2230 is estimated
to be $1.8 \times 10^8$ G [1], though the magnetic field inside the star
may be larger.

Recently, strong magnetic fields, $B \sim 10^{17}$ G, in a proto-neutron star
were introduced to explain the pulsar kick in Ref. \cite{Maruyama:2010hu}.
With such strong magnetic fields and propagation of neutrinos in a proto-neutron star,
kick velocities of a few hundred km per second are calculated.
Therefore, consideration of magnetic fields can open
various possibilities by which large neutron masses and pulsar kicks can be explained.

\section*{ACKNOWLEDGMENTS}
This work was supported in part by the Korea Research Foundation Grant
funded by the Korean Government (KRF-2008-313-C00208) and by the WCU program
(R31-2008-000-10029-0) funded by the Ministry of Education, Science and Technology.

\end{document}